\documentclass[aps,prc,twocolumn,superscriptaddress]{revtex4}
\usepackage{graphicx}
\usepackage{dcolumn}
\usepackage{bm}
\usepackage{siunitx}
\newcommand{\tol}[3]{\ensuremath{\si{#1}^{\thinspace+\num{#2}}_{-\num{#3}}}} 

\begin{document}
\title{New limit of  $^{244}$Pu on Earth points to rarity of actinide nucleosynthesis}

\author{A. Wallner} \email[corresponding author: ]{anton.wallner@anu.edu.au}
\affiliation{Department of Nuclear Physics, Research School of Physics and Engineering, The Australian National University, Canberra, Australia} 
\affiliation{VERA Laboratory, Faculty of Physics, University of Vienna, W{\"a}hringer Strasse 17, A-1090, Austria} 
\author{T. Faestermann} \affiliation{Technische Universit{\"a}t M{\"u}nchen, D-85747 Garching, Germany}
\author{J. Feige} \affiliation{VERA Laboratory, Faculty of Physics, University of Vienna, W{\"a}hringer Strasse 17, A-1090, Austria}
\author{C. Feldstein} \affiliation{Racah Institute of Physics, Hebrew University, Jerusalem, Israel}
\author{K. Knie} \affiliation{Technische Universit{\"a}t M{\"u}nchen, D-85747 Garching, Germany}  \affiliation{GSI Helmholtzzentrum f{\"u}r Schwerionenforschung GmbH, Planckstr. 1, 64291 Darmstadt, Germany}
\author{G. Korschinek} \affiliation{Technische Universit{\"a}t M{\"u}nchen, D-85747 Garching, Germany}
\author{W. Kutschera} \affiliation{VERA Laboratory, Faculty of Physics, University of Vienna, W{\"a}hringer Strasse 17, A-1090, Austria}
\author{A. Ofan} \affiliation{Racah Institute of Physics, Hebrew University, Jerusalem, Israel}
\author{M. Paul} \affiliation{Racah Institute of Physics, Hebrew University, Jerusalem, Israel}
\author{F. Quinto}  \altaffiliation[now: ] {Institute for Nuclear Waste Disposal, Karlsruhe Institute of Technology (KIT), Campus North, Karlsruhe, Germany} \affiliation{VERA Laboratory, Faculty of Physics, University of Vienna, W{\"a}hringer Strasse 17, A-1090, Austria}
\author{G. Rugel} \altaffiliation[now: ] {Helmholtz-Zentrum Dresden-Rossendorf, Bautzner Landstraße 400 01328 Dresden, Germany} \affiliation{Technische Universit{\"a}t M{\"u}nchen, D-85747 Garching, Germany}
\author{P. Steier} \affiliation{VERA Laboratory, Faculty of Physics, University of Vienna, W{\"a}hringer Strasse 17, A-1090, Austria}

\begin{abstract}
Half of the heavy elements including all actinides are produced in r-process nucleosynthesis whose sites and history still remain a mystery. If continuously produced, the Interstellar Medium (ISM) is expected to build up a quasi-steady state of abundances of short-lived nuclides (with half-lives $\leq{100}$ My), including actinides produced in r-process nucleosynthesis. Their existence in today’s ISM would serve as a radioactive clock and would establish that their production was recent. In particular $^{244}$Pu, a radioactive actinide nuclide (81 My half-life), can place strong constraints on recent r-process frequency and production yield. Here we report on the detection of live interstellar $^{244}$Pu, archived in Earth’s deep-sea floor during the last 25 My, at abundances lower by about two orders of magnitude than expected from continuous production in the Galaxy. This large discrepancy may signal a rarity of actinide r-process nucleosynthesis sites, compatible with neutron-star mergers or with a small subset of actinide-producing supernovae.

\end{abstract}
\maketitle

\section{Introduction \label{sec:1}}

About half of all nuclides existing in nature and heavier than iron are generated in stellar explosive environments. Their production requires a very short and intense burst of neutrons (rapid neutron-capture or r-process) \cite{Qia03,WW95,AGT07}. The nuclides are formed via successive neutron-captures on seed elements, following a path in the very neutron-rich region of nuclei. However, the relevant astrophysical sites, with supernovae (SN) \cite{Qia03,WW95} and neutron-star (NS-NS) mergers \cite{AGT07,Thi11} as prime candidates, and the history of the r-process during the Galactic chemical evolution are largely unknown. The Interstellar Medium (ISM) is expected to become steadily enriched with fresh nucleosynthetic products and may also contain continuously produced short-lived nuclides (with half-lives $\leq{100}$ My) \cite{MC00}, including actinides produced in r-process nucleosynthesis.

Recent r-process models within SNe-II explosions, based on neutrino wind scenarios \cite{AMP11,Gor13}, suffer difficulties on whether heavy elements can really be produced in these explosions. An alternative site is neutron-star ejecta, e.g. NS-NS, or neutron-star black-hole (NS-BH) mergers. Candidates of such neutron star binary systems have been detected \cite{Tan13}. Estimations of an NS-NS merger event rate of about (2-3)$\times$10$^{-5}$ per year in our galaxy would allow for such mergers to account for all heavy r-process matter in our Galaxy \cite{AGT07,AST04,GA01}. 

As pointed out by Thielemann et al. \cite{Thi11} "Observations of low metallicity stars indicate also here the probable splitting in more types of events: (a) a rare event, reproducing the heavy r-process abundances always in solar proportions, and (b) a more frequent event, responsible for the lighter r-process abundances". Galactic Chemical Evolution models \cite{JF13, AST04,Thi02} show that neutron-star mergers, occurring at late time in the life of a galaxy, cannot account for all the r-process nuclei found in very old stars \cite{JF13}. Thus, recent models suggest different r-process scenarios (similar to s process) which might occur at different nucleosynthesis sites \cite{AGT07,Thi02}.

To summarize, very few hints on astrophysical sites and galactic chemical evolution exist: (i) the relative abundance distribution observed spectroscopically in a few old stars for r-process elements between barium and hafnium is very similar to that of the Solar System (SS) \cite{CS06,Qia03}, pointing to an apparently robust phenomenon; a large scatter for the r-process elements beyond Hf and also below barium is however observed \cite{AGT07,GA01,FTC02}; (ii) the Early Solar System (ESS) is known to have hosted a set of short-lived radioactive nuclides (t$_{1/2}$ $\lesssim$100 My) \cite{MC00,HMS09,Die06,WBG06}, among them pure r-process nuclei such as $^{244}$Pu (81 My half-life) and  $^{247}$Cm (15.6 My), clearly produced no more than a few half-lives before the gravitational collapse of the protosolar nebula \cite{THH07,Kur60,Lac12,HLM71}.

We report here on a search for live $^{244}$Pu (whose abundance in the ESS relative to $^{238}$U was ca. 0.8 \% \cite{MC00,THH07,Kur60,HLM71,Dau05,HKP89} in deep-sea reservoirs, which are expected to accumulate Interstellar Medium (ISM) dust-particles over long time periods. 

Our findings indicate that SNe, at their standard rate of $\sim${1-2/100} y in the Galaxy, did not contribute significantly to actinide nucleosynthesis for the past few hundred million years. A similar conclusion is drawn, when related to the recent SNe history in the LB: we do not find evidence for live $^{244}$Pu that may be locked in the ISM in accumulated swept-up material and that was transported to Earth by means of the recent LB SNe activity. Our results suggest that actinide nucleosynthesis, as mapped through live $^{244}$Pu, seems to be very rare.

\section{Results: \label{sec:2}}

\subsection{Experimental concept}

ISM dust particles \cite{Dwe98,Man10}, assumed to be representative of the ISM, are known to enter the SS and are expected to reach and accumulate on Earth in long-term natural depositories such as deep-sea hydrogenous iron-manganese (FeMn) encrustations and sediments. Its detection would be the equivalent for r-process nuclides of the $\gamma$-ray astronomy observations of live radioactivities \cite{Die06} produced by explosive nucleosynthesis in single SN-events (e.g. $^{56}$Ni (6.1 d), $^{56}$Co (77.3 d), $^{44}$Ti (60.0 y) or diffuse in the Galactic plane such as $^{26}$Al (0.72 My) and  $^{60}$Fe (2.62 My), owing to their longer half-life. 

Several models, based on the frequency of SN-events, the nucleosynthesis yield and the radioactive half-life, were developed to calculate the abundance of a nucleosynthesis product (t$_{1/2}$ $>$10 My) in quasi-secular equilibrium between production and radioactive decay-rates. These models together with the flux and average mass of ISM dust-particles into the inner SS measured by space missions in the last decade (Galileo, Ulysses, Cassini) \cite{Alt05} are used here to estimate the corresponding influx of $^{244}$Pu nuclei onto Earth.

We compare our results also with a possible imprint of recent actinide nucleosynthesis ($<$15 My) from the SNe history of the Local Bubble (LB, a cavity of low density and hot temperature of $\sim$200 pc diameter). Recent ISM simulations suggest about 14-20 SN explosions within the last 14 My \cite{AL02,FBA06,BB14} that were responsible for forming the local ISM structure and the LB. $^{244}$Pu decay can be considered negligible during this period. The SN ejecta shaped the ISM and accumulate also swept-up material including pre-existing $^{244}$Pu from nucleosynthesis events prior to the formation of the LB \cite{EFS96,FHE05}.

With a growth-rate of a few mm/My \cite{Seg84}, hydrogenous crusts will strongly concentrate elements and particles present in the water column above. The higher accumulation rate of deep-sea sediments (mm/ky) results in a better time resolution but requires much larger sample volumes. With regard to other potential $^{244}$Pu sources, we note that natural $^{244}$Pu production on Earth is negligible and the ESS abundance has decayed to 10$^{-17}$ of its pre-solar value \cite{Lac12,HLM71}.

Anthropogenic production from atmospheric nuclear bomb-tests and from high-power reactors is restricted to the last few decades, localized in upper layers and can easily be monitored through the characteristic isotopic fingerprint of the other co-produced $^{239-242}$Pu isotopes. In fact the detection of anthropogenic $^{239,240}$Pu in deep-sea sediments \cite{LA83,BNL80,Pau01} and crusts \cite{Wal04} provides an excellent proxy for the ingestion efficiency of dust from the high atmosphere into these reservoirs, together with their chemical processing towards the final analyzed samples (see Methods).

\subsection{Selected terrestrial archives for extraterrestrial $^{244}$Pu}

Terrestrial archives, like deep-sea iron-manganese crust and sediment archives extend over the past tens of million years. Large dust grains entering Earth's atmosphere have also been observed by radar detections \cite{Bag00}. Extraterrestrial dust particles, cosmogenic nuclides and terrestrial input sink to the ocean floor and are eventually incorporated into the FeMn-crust or sediment. Such a process is confirmed by inclusion in these archives of meteoritic  $^{10}$Be, cosmogenic  $^{53}$Mn and live  $^{60}$Fe, the latter attributed to the direct ejecta of a close-by SN \cite{Kni04,Fit07,Pou07}. For actinide transport through the latter stages, the observed deposition of global fallout \cite{Wal04} from atmospheric nuclear bomb testing \cite{LA83,BNL80} in deep marine reservoirs after injection to the stratosphere serves as a proxy to extraterrestrial particles. 

We chose two independent archives: a large piece (1.9 kg and 0.4 kg samples) from a deep-sea manganese crust (237KD from cruise VA13/2, collected in 1976) with a growth-rate between 2.5 mm  My$^{-1}$ \cite{Fit07,Seg84} and 3.57 mm  My$^{-1}$ \cite{SNB03}. It originates from the equatorial Pacific (location  $9^\circ$ $180^\prime$ N,  $146^\circ$ $030^\prime$ W) at a depth of 4830 m and covers the last  $\sim$25 million years \cite{Pou07}. In the very same crust, the live  $^{60}$Fe signal mentioned above was found at about 2.2 My before present (BP) \cite{Kni04,Fit07}. Our second sample, also from the Pacific Ocean, is a piston core deep-sea sediment (7P), extracted during the TRIPOD expedition as part of the Deep-Sea Drilling Project (DSDP) at location $17^\circ$ $30^\prime$ N, $113^\circ$ $00^\prime$ W at 3763 m water depth and covers a time period of $\sim$1.64 My (0.53-2.17 My BP) \cite{Smi14}. 

The crust sample, covering a total area of 227.5 cm$^{2}$ and a time range of 25 My, was split into four layers (1-4) representing different time periods in the past (see Table I). Each layer was subdivided into three vertical sections (B, C and D) with areas between 70 and 85 cm$^{2}$, totaling 12 individually processed samples. The surface layer (layer 1, with a time-range from present to 500,000 years BP) contains also the anthropogenic Pu signal originating from global fallout of atmospheric weapons testing \cite{LA83,BNL80}. The next layer 2 spans a time period from 0.5--5 My BP, layer 3 5--12 My and layer 4 12--25 My \cite{Pou07} (we note the age for samples older than 14 My, where no  $^{10}$Be dating is possible \cite{Fit07,Seg84}, is more difficult to establish; different age models suggests a time period of 12--$\sim$18-20 My \cite{FNH99}, another model up to  $\sim$30 My \cite{SNB03} for layer 4). Finally, sample X, the bottom layer of hydrothermal origin (Fig. 1) served as background sample. 

For archives accumulating millions of years, the expected $^{244}$Pu abundance range (see discussion) is well within reach of accelerator mass spectrometry (AMS), an ultra-sensitive method \cite{Ste13,Syn13,Kut13} of ion identification and detection. Based on the ingestion efficiency of Pu into deep-sea manganese crusts (21\%) and on the AMS $^{244}$Pu detection efficiency ($1\times10^{-4}$, see Methods and Table III), we calculate a measurement sensitivity expressed as a $^{244}$Pu flux onto Earth of the order of ten atoms cm$^{-2}$ My$^{-1}$ $^{244}$Pu from ISM deposition; i.e. for the crust with a 25 My accumulation period and with 200 cm$^{2}$ surface area  $\sim$500-4000 $^{244}$Pu detection events are expected; and about a factor 100 less for the sediment sample (1.64 My time period and 4.9 cm$^{2}$ surface area).

\vspace*{2mm}

\onecolumngrid
\begin{center}
\begin{figure}[ht]
\fboxsep=0pt
\includegraphics[width=14. cm]{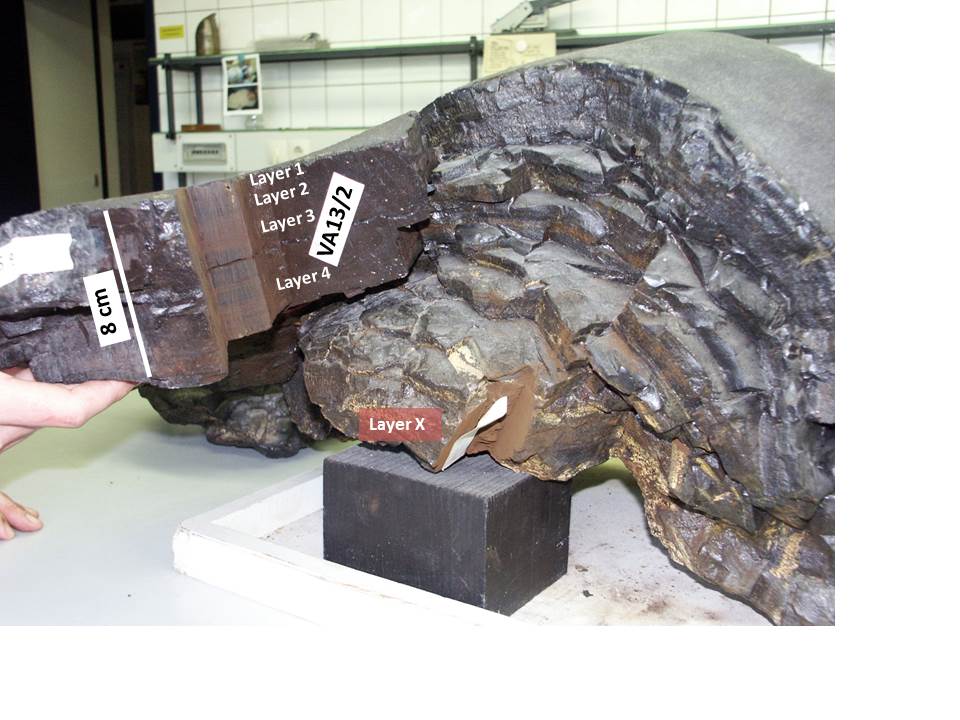}
\caption{(Color online) Picture of the crust sample 237KD -- this FeMn crust sample was sampled 
1976 from the Pacific Ocean at 4830 m water depth. Large samples (with a total thickness of 25 cm) used in this work were taken from one part of the crust (hydrogenous crust, layers 1 to 4, left in the figure) and from the bottom (hydrothermal origin, layer X, crust started to grow  $\sim$65 My ago \cite{SNB03}, see also Frank et al. \cite{FNH99})}.
\label{fig:KD237}
\end{figure}
\end{center}
\twocolumngrid


\subsection{AMS experimental data of $^{244}$Pu abundances in Earth archives}

We have developed the capability to detect trace amounts of $^{244}$Pu in terrestrial archives by AMS \cite{Ste13} and our technique provides background-free $^{244}$Pu detection with an overall efficiency (atoms detected/atoms in the sample) of $\sim$1$\times{10^{-4}}$ (Table III and supplementary information). The AMS measurements determine the atom ratio $^{244}$Pu/$^{A}$Pu where $^{A}$Pu (A= 236 or 242) is a spike of known amount (added during the chemical processing of the sample) from which the number of $^{244}$Pu nuclei in the sample is obtained (see Methods, Table III and supplementary information). In addition to $^{244}$Pu counting, we also measured the shorter-lived $^{239}$Pu (t$_{1/2}$ = 24.1 ky) content as an indicator of anthropogenic Pu signature.

The results for the four crust layers and the blank sample, obtained from the AMS measurements on 11 individual crust samples, are listed in Table I and III (see supplement with details; identification spectra are plotted in Figure 2). We observed one single event in each of two crust subsamples, namely layer 3, section B (B3) and layer 4, section D (D4). No $^{244}$Pu was registered in the other 7 crust subsamples or the blank sample (X). 

A clear anthropogenic $^{239}$Pu and $^{244}$Pu signal was observed in the top layer (16 events of $^{244}$Pu detected). Measurements of samples from deep layers ($>$0.5 My) show also some events during the $^{239}$Pu measurement (compared to the top layer, the $^{239}$Pu countrate in the deep crust layers were a factor of  $\sim$100 lower, and the one in the sediment and blank sample a factor of  $\sim$1000 lower). Since naturally produced $^{239}$Pu in these older layers is considered negligible, its presence is attributed to $^{238}$U still present in the final AMS sample at about 8 to 9 orders of magnitude higher than $^{239}$Pu and mimicking $^{239}$Pu detector events (see Methods); we also note that the $^{236}$Pu spike added for tracing the measurement efficiency was found to contain some $^{239}$Pu which we corrected for, see supplementary table T4). 

\onecolumngrid
\begin{center}
\begin{figure}[ht]
\fboxsep=0pt
\includegraphics[width=14. cm]{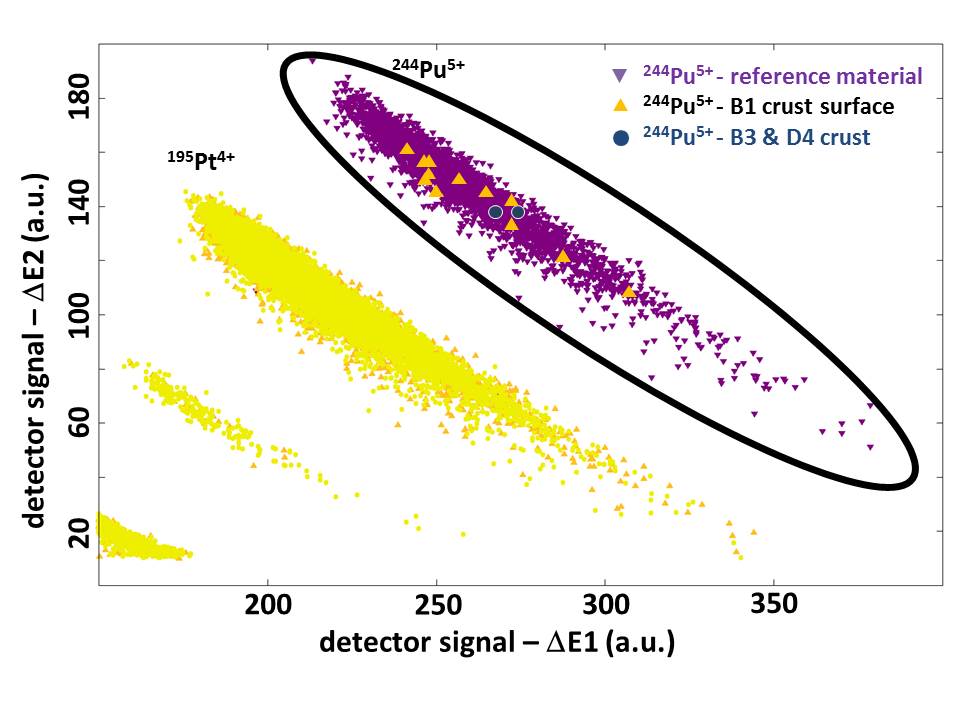}
\caption{Identification spectra obtained in the AMS measurements with a particle detector (two independent differential energy-loss signals ($\Delta$E1, $\Delta$E2) are plotted in x- and y-axis). Parasitic (or background) particles of different energy (e.g. $^{195}$Pt$^{4+}$) and different mass were clearly separated and do not interfere. Displayed is an overlay of $^{244}$Pu$^{5+}$ events obtained in a series of measurements for a $^{244}$Pu reference material (purple triangle), the 12 events registered for one of the surface layer samples, B1 (yellow triangle), and the 2 events measured for the deeper layers B3 and D4, respectively (blue circle).}.
\label{fig:spectra}
\end{figure}
\end{center}
\twocolumngrid

We conclude from the $^{239}$Pu data that anthropogenic contamination did not add any significant $^{244}$Pu detector events for all older layers (using the anthropogenic $^{244}$Pu/$^{239}$Pu ratio obtained from the top layer, $1\times 10^{-4}$, see Methods, Table III). In the following we calculate for all cases 2$\sigma$ upper limits, i.e. 95\% confidence levels and 0 (1 or 2) $^{244}$Pu events corresponds to an upper limit of $<$3 (5 or 6.7, respectively) $^{244}$Pu events (applying statistics for small signals \cite{FC98}).

\begin{table*}[bt]
\caption{\label{tab_1}$^{244}$Pu detector events and corresponding ISM flux compared to galactic chemical models assuming steady-state}
\begin{ruledtabular}
\begin{tabular}{lccccccc}
deep- sea 						&time				& sample 		& sample 		& integral sensitivity 	& $^{244}$Pu detector  	& $^{244}$Pu flux into 		& $^{244}$Pu flux ISM at 		\\
archive						&  period			& area		& mass		& (eff.$\times$area$\times$time 	& events		& terrestrial archive 	&  Earth orbit 				\\[1ex]
							& (My)			& (cm$^{2}$)	& (g)			& period) (cm$^{2}$$\cdot$My)	 	& (2$\sigma$ limit)$^{1}$		& (at$\cdot$cm$^{-2}$$\cdot$My$^{-1}$)		& (at$\cdot$cm$^{-2}$$\cdot$My$^{-1}$)$^{2}$				\\
\hline \\
\vspace*{1mm}crust{\_}modern					&0--0.5			& 227.2		& 80			&	0.006			& 16			&  --			&  --						\\[1ex]
\vspace*{1mm}Layer X						&blank			& ${\sim}$100	& 364			&	--			&   0			&  --			&  --						\\[1ex]
\hline \\
Layer 2						&0.5--5			& 227.2		& 473			&	0.016			& 0 ($<$3)		& $<$188				& $<$3500					\\[1ex]
Layer 3						&5--12				& 227.2		& 822			&	0.075			& 1 ($<$5)	& \tol{13}{53}{12} ($<$66)	& \tol{247}{1000}{235}			\\[1ex]
Layer 4						&12--25			& 142.2		& 614			&	0.060			& 1 ($<$5)		& \tol{17}{66}{16} ($<$83)	& \tol{320}{1250}{300}			\\[1ex]
\hline \\
\bf{Crust	(2--4)}						&0.5--25			& 182			& 1909	& 0.151		& 2 ($<$6.7)		& \tol{13}{31}{11} ($<$44)	& {\bf{\tol{250}{590}{205} } }		\\[1ex]
\hline \\
\bf{Sediment}						&0.53--2.17			& 4.9			& 101 			& 0.0013		& 1 ($<$5)		& \tol{750}{3000}{710}		& {\bf{\tol{3000}{12000}{2850}} }	\\[1ex]	
\hline \\
\multicolumn{7}{l}{ {\bf{Model}} --  steady-state model \& ISM flux data at 1 AU from satellite Cassini$^3$}													& \bf{20000--160000}			\\[1ex]	
\end{tabular}
\end{ruledtabular}

\begin{flushleft}
The FeMn crust sample was split into four layers 1--4 and three sections B, C and D. The top layer (1 mm, 'crust modern') was removed for measuring the anthropogenic Pu content originating from atmospheric atomic-bomb tests from  $\sim$1950 to 1963. In total 2 $^{244}$Pu detector events were registered using AMS in all older crust samples over a 72 h counting time (column 6). We calculate from our data an extraterrestrial $^{244}$Pu flux and a 2$\sigma$-limit of $<$6.7 extraterrestrial $^{244}$Pu events \cite{FC98}. The sediment sample also gave one $^{244}$Pu detector event and none were registered in any of the blank samples. The term 'integral sensitivity' represents a quantity that combines the overall measurement efficiency (eff.), the flux integration area and the time period covered by the individual samples. \\
\vspace*{1mm}
$^{1}$ because of the low $^{244}$Pu event rate, we also display 2$\sigma$ upper levels (95\% confidence levels) applying low-level statistics \cite{FC98}. \\
\vspace*{1mm}
$^{2}$ using an incorporation efficiency $\epsilon$=(21$\pm$5)\% for the crust and 100\% for the sediment sample (Methods). The mean area for the crust sample is 182 cm$^{2}$ (accounting for the different time periods) and 4.9 cm$^{2}$ for the sediment sample. For calculating the ISM flux at Earth orbit, the measured $^{244}$Pu flux into the terrestrial archives was corrected for the incorporation efficiency and was multiplied by a factor of 4 to account for the ratio of Earth's surface to its cross section (i.e. assuming an unidirectional and homogeneous ISM flux relative to the solar system). \\
\vspace*{1mm}
$^{3}$see Table II: the steady-state $^{244}$Pu flux is based on the actinide (U and Th) abundances measured in meteorites, and on present-day Pu/U and Pu/Th ISM concentrations deduced from two independent models for galactic chemical evolution: (1) uniform production model by Wasserburg et al. \cite{WBG06} and (2) the model developed by Clayton et al. \cite{Cla11,MC00,Dau05}. The Pu flux at 1 AU (Earth orbit) is normalized to the measured interstellar dust particles (ISD) flux deduced from the satellite (Cassini) data \cite{Alt05}, i.e. accounts for the filtration of ISD when entering the heliosphere of our solar system (3-9\%, see Methods). \\
\vspace*{1mm}
\end{flushleft}	
\end{table*}

\subsection{$^{244}$Pu flux deduced from measured terrestrial concentrations}

The crust data for all sections and for all three deeper layers are compatible (for details see also Supplementary Table 1). Owing to a higher chemical yield (integral sensitivity, col. 5, Table I) layers 3 and 4 provide lower limits. The measured $^{244}$Pu concentration into these layers can be converted into a $^{244}$Pu particle flux using chemical yield, detection efficiency, the incorporation efficiency of Pu into the crust (21$\pm$5 \%, see Methods and Table III), and the area and time period covered. We also assume the extraterrestrial $^{244}$Pu flux through Earth’s cross-section is homogenously distributed over the Earth’s surface. Hence, the interstellar flux is calculated by multiplying the measured flux into the crust by a factor of 4/0.21 = 19. We derive thus a 2$\sigma$ limit \cite{FC98} for the $^{244}$Pu-ISM-flux at Earth orbit from data of the three layers $<$3,500, $<$1,300 and $<$1,560 $^{244}$Pu ats cm$^{-2}$ My$^{-1}$ and the one $^{244}$Pu event in layers 3 and 4 corresponds to a flux of \tol{247}{1000}{235} and  \tol{320}{1250}{300} ats cm$^{-2}$ My$^{-1}$, respectively. Combining all samples (2 $^{244}$Pu events) a flux of \tol{250}{590}{205} and a 2$\sigma$-limit on the $^{244}$Pu-flux of $<$840 $^{244}$Pu ats cm$^{-2}$ My$^{-1}$ is obtained. The data are plotted in Figure 2. The single $^{244}$Pu event measured for the deep-sea sediment converts to a flux of 3,000 ($<$15,000) ats cm$^{-2}$ My$^{-1}$. Both archives give consistent $^{244}$Pu flux limits with a higher sensitivity for the crust samples.

\begin{table*}[bt]
\caption{\label{tab.2} Steady-state concentrations of $^{244}$Pu from galactic chemical evolution models; Cassini satellite measurements \cite{Alt05,Alt03} of the interstellar particle flux at 1 AU (Earth orbit) and the deduced $^{244}$Pu fluxes at 1 AU.}
\begin{ruledtabular}
\begin{tabular}{lcc} \\
\multicolumn{3}{c}{\bf{meteoritic U \& Th abundance data \& models for the present Pu/U and Pu/Th ISM ratio}}														\\
\hline \\
\vspace*{1mm}dust mass density local galactic environment (g cm$^{-3}$) \cite{Man10,Li05}		& \multicolumn{2}{c}{1.2$\times$10$^{-26}$}								 \\
\vspace*{1mm}U abundance in dust (from meteorite data) (g/g)  							& \multicolumn{2}{c}{17$\times$10$^{-9}$ g U/ g meteorite$^{B}$} 					\\
\vspace*{1mm}Pu abundance in Early SS (measured from fissiogenic Xe)	 					& \multicolumn{2}{c}{$^{244}$Pu/$^{238}$U=0.008 at/at \cite{THH07,Dau05}} 					\\
\hline \\
galactic chemical														&Uniform Production (UP) 				& Open-box model  \\
\vspace*{2mm} evolution model												&Wasserburg et al. \cite{WBG06}		& \cite{Cla11,MC00,Dau05,Dau05b}		\\

model prediction for present ISM steady-state 									&  							& 								\\
\vspace*{2mm}ratio $^{244}$Pu/$^{238}$U (at/at):								& 0.0165						& 0.044							\\

Present expected $^{244}$Pu concentration in ISM dust  								&  							& 			\\
\vspace*{2mm}(using meteoritic data as proxy)									&2.8$\times$10$^{-10}$ g/g				& 7.5$\times$10$^{-10}$ g/g			\\

\vspace*{2mm}$^{244}$Pu ISM concentration									&0.8$\times$10$^{-14}$ at/cm$^{3}$		& 2.2$\times$10$^{-14}$ at  cm$^{-3}$			\\

\vspace*{1mm}range of  $^{244}$Pu ISM concentration 								& \multicolumn{2}{c}{\bf{(0.8--2.2)$\times$10$^{-14}$  $^{244}$Pu at  cm$^{-3}$}}  	\\
\hline \\
\vspace*{1mm}Cassini: flux data at Earth orbit \cite{Alt05} 								& \multicolumn{2}{c}{(3-4)$\times$10$^{-5}$ particles m$^{-2}$ s$^{-1}$}  	\\  
\vspace*{1mm}Cassini data: mean particle mass at Earth orbit \cite{Alt05,Alt03}				& \multicolumn{2}{c}{(3-7)$\times$10$^{-13}$ g (0.5--0.6 $\mu$m radius, $\rho$=2.5 g  cm$^{-3}$)}  	\\  
\vspace*{1mm}Cassini: mean mass flux											& \multicolumn{2}{c}{(9-28)$\times$10$^{-22}$ ISM g cm$^{-2}$ s$^{-1}$}  	\\  
\vspace*{1mm}Cassini mass flux $\times$  $^{244}$Pu conc. in ISM dust 					& \multicolumn{2}{c}{(2.5-20)$\times$10$^{-31}$  $^{244}$Pu g cm$^{-2}$ s$^{-1}$}  	\\  
\hline \\
 $^{244}$Pu flux at Earth orbit												& \multicolumn{2}{c}{\bf{20,000--160,000  $^{244}$Pu at cm$^{-2}$ My$^{-1}$ } } 	\\  
 \\
fraction of ISM dust found at 1 AU											&  							& 								\\
(using a peak velocity of 26 km s$^{-1}$)										& \multicolumn{2}{c}{3--9 \%}  	\\  
\hline \\
\multicolumn{3}{c}{\bf{$^{244}$Pu fluence through a surface of 75 pc radius in swept-up material of the Local Bubble} }														\\
\hline \\
pre-LB intermediate dust mass density (required to form 								&	  						& 								\\
\vspace*{2mm}the Local Bubble, 7 particles  cm$^{-3}$ \cite{BB14} with 1\% of mass in dust)		& \multicolumn{2}{c}{7.6$\times$10$^{-26}$ g  cm$^{-3}$}  	\\  
\hline \\

\vspace*{1mm}ISM dust column density over 75 pc (radius of LB) 							& \multicolumn{2}{c}{2$\times$10$^{-5}$ g cm$^{-2}$ }  	\\  

total  $^{244}$Pu fluence from swept-up ISM 										&  							& 								\\
\vspace*{2mm}dust into SS at Earth orbit 										& \multicolumn{2}{c}{\bf{(0.4--3)$\times$10$^{6}$  $^{244}$Pu at cm$^{-2}$} } 	\\  

\end{tabular}
\end{ruledtabular}

\begin{flushleft}
1 solar mass = 1.99$\times$10$^{30}$ kg 	- 	1 parsec (pc) = 30.857$\times$10$^{15}$ m (1 pc$^{3}$=2.94$\times$10$^{55}$ cm$^{3}$) \\
$^{B}$ The measured U concentration in present meteorites of (8.4${\pm}$0.8) ppb (Lodders 2003)\cite{Lod03} was adjusted for decay ($\sim$one half-life of $^{238}$U). 

\end{flushleft}	

\end{table*}

\section{Discussion: \label{sec:3}}

\begin{table*}[bt]
\caption{\label{tab_3}Anthropogenic Pu at the surface and incorporation efficiency (eff.) into the manganese crust}
\begin{ruledtabular}
\begin{tabular}{lccccc}
 						&\multicolumn{4}{c}{Surface Layer 1}					& Blank			\\[1ex]
Time period					&\multicolumn{4}{c}{0--0.5 My}						& --		         		\\[1ex]
						&\multicolumn{4}{c}{contains anthropogenic Pu (top 1 mm)}	& hydrothermal		\\
						&\multicolumn{4}{c}{ }							& $\sim$100 cm$^{2}$    	\\[1ex]
\hline \\
subsample					&B1			& C1		& D1			&	total			& X		\\[1ex]
mass (g)					&32			& 20		& 28			&	80			&364		\\[1ex]
total meas. eff. (10$^{-4}$)		&0.82			& 0.45	& 0.18		&	$<$0.51$>$		&0.93		\\[1ex]
measuring time				&3.8 h		& 3.8 h	& 2.6h		&	10.2 h			&3.4 h	\\[1ex] 
$^{236}$Pu atoms spike			&3$\times$10$^{8}$		&3$\times$10$^{8}$	&3$\times$10$^{8}$			&9$\times$10$^{8}$			&3$\times$10$^{8}$		\\[1ex]
$^{244}$Pu atoms (10$^{4}$)		&18.5			& 5.3		& 13.3		&	37.1			&$<$1.7	\\[1ex]
$^{239}$Pu atoms (10$^{8}$)		&13.9			& 15.6	&9.9			&	39.4			&0.3		\\[1ex]
$^{244}$Pu/$^{239}$Pu 			&1.3			& 0.3		& 1.3			&	1.0$\pm$0.3		&--		\\[1ex]

$^{239}$Pu atoms cm$^{-2}$ measured		&1.6$\times$10$^{7}$			& 2.2$\times$10$^{7}$		& 1.4$\times$10$^{7}$			&	1.76$\times$10$^{7}$			& 		\\[1ex]
$^{239}$Pu atoms cm$^{-2}$ reaching &			& 		& 			&					& 		\\
deep-sea floor$^{1}$ in 1976 \cite{LA83,BNL80}			&			& 		& 			&	8.2$\times$10$^{7}$	& 		\\[1ex]
$^{239}$Pu incorporation eff. (crust)	&			& 		& 			&	21$\pm$5 \%		& 		\\[1ex]
\end{tabular}
\end{ruledtabular}

\begin{flushleft}
237KD (VA13/2) deep-sea crust measurement: detailed data for the surface layer 1 (anthropogenic Pu) and the hydrothermal blank sample and determination of the Pu incorporation efficiency (eff.) into the deep-sea manganese crust by comparison of the known amount of atomic-bomb produced Pu at the crust’s location with the measured Pu in the top layer 1.

\vspace*{1mm}
$^{1}$ The amount of $^{239}$Pu ats cm$^{-2}$ reaching the deep-sea floor at the time of crust sampling (1976) is derived from the ratio (2.1 \%) of the $^{239,240}$Pu fluence measured in deep-sea sediments \cite{LA83} (assumed to incorporate 100\% of precipitated material) and the overall $^{239,240}$Pu fallout fluence measured for the location of the crust \cite{BNL80}.

\vspace*{1mm}
The total measurement efficiency (eff.) in our work is calculated from the total number of $^{236}$Pu registered, normalized by the time fraction of $^{236}$Pu AMS counting and divided by the number of $^{236}$Pu atoms added as spike to the sample (3$\times$10$^{8}$ atoms each). 

\vspace*{1mm}
The $^{244}$Pu atoms per sample are calculated from the number of $^{244}$Pu events registered with the particle detector, scaled by the time fraction of AMS $^{244}$Pu counting and normalized with the measurement efficiency; ditto for $^{239}$Pu atoms per sample. The $^{239}$Pu detector events were corrected for a well-known contribution when adding the $^{236}$Pu spike which contains also $^{239}$Pu (see Supplementary Table 1 for more details).

\end{flushleft}	
\end{table*}

First, we estimate the expected $^{244}$Pu flux from ISM dust particles penetrating the solar system and their incorporation into terrestrial archives. Our experimental results are then compared with these estimations. Based on a Uniform Production model (UP) cite{WBG06} or an open-box model \cite{Cla11,MC00,HMS09} (see also \cite{Dau05}) taking into account Galactic-disk enrichment in low-metallicity gas, the present day ISM atom-ratio $^{244}$Pu/$^{238}$U from SN events is calculated to be between 0.017 and 0.044 (Table II). We further assume that the abundance of $^{238}$U in ISM dust is the same as that of chondrite meteorites \cite{Lod03} (corrected for the SS-age), 1.7$\times$10$^{-8}$ g/g, and derive a steady-state $^{244}$Pu abundance of (2.8--7.5)$\times$10$^{-10}$ g Pu/g ISM (i.e. (0.8--2.2)$\times$10$^{-14}$ at cm$^{-3}$; similar values are obtained if normalized to $^{232}$Th. 

For interstellar dust particles (ISD) entering the solar system \cite{Man10}, we have to take into account filtering when penetrating the heliosphere. ISD were observed by the Ulysses, Galileo and Cassini \cite{Alt05,Alt03} space missions over more than 5 years, for distances from the Sun between 0.4 and $>$5 AU. Measurements of the Cassini space mission \cite{Alt05,Man10} determine a mean flux of ISM dust of (3--4)$\times$10$^{-5}$ particles m$^{-2}$ s$^{-1}$ at a distance of 1 AU, i.e. at Earth's position, with a mean particle mass of (3--7)$\times$10$^{-13}$ g (0.5--0.6 $\mu$m average particle size). These particles show a speed distribution corresponding to the flow velocity of the ISM (26 km s$^{-1}$) and constitute 3-9\% of the dust component of the ISM intercepted by the SS (see Table II). The direct collection of a few particles identified as ISD, very recently reported \cite{Wes14}, although yet of low statistical significance, supports the scenario of penetration of large ISD particles into the inner SS and may be consistent with the satellite data. It should be noted that Galactic cosmic-rays penetrate the SS and recent observations clearly demonstrate therein the presence of Th and U, and tentatively of $^{244}$Pu \cite{Don12}.

Within the assumptions described above, the expected flux of $^{244}$Pu atoms from the ISM reaching the inner SS (at Earth orbit) is (2.5--21)$\times$10$^{-31}$ g Pu cm$^{-2}$ My$^{-1}$ or 20,000--160,000 $^{244}$Pu atoms cm$^{-2}$ My$^{-1}$. If evenly distributed over Earth’s surface (i.e. assuming a unidirectional ISM flux) the $^{244}$Pu flux into terrestrial archives becomes 5,000--40,000 $^{244}$Pu atoms cm$^{-2}$ My$^{-1}$.

Our experimental results (Table I) provide for the first time a sensitive limit of interstellar $^{244}$Pu concentrations reaching Earth, integrated over a period of 24.5 My. Our data are a factor of 80--640 lower than the values expected under our constraints on ISM grain composition from a SN derived steady-state actinide production (the 2$\sigma$ upper limit of  $\sim$840 ats cm$^{-2}$ My$^{-1}$ is still a factor of 25--200 lower). The lifetime of $^{244}$Pu is comparable to the complete mixing time scales of the ISM \cite{AL02,FBA06,BB14}. The deep-sea crust sample integrates a $^{244}$Pu flux over a time period of 24.5 My ($\sim$1/10 of the SS rotation period in the Galaxy) corresponding to a relative travel distance of the SS of 650 pc (taking the mean speed of the measured ISM dust particles of 26 km/s \cite{Alt03},  $\sim$1/10 of the galactic orbital speed, as a proxy for ISM reshaping and for motion differences relative to the co-rotating local neighborhood). These results, consistent with previous studies on extraterrestrial $^{244}$Pu in crust \cite{Wal04,WFG00} and sediment samples \cite{Pau01,Pau07}, are a factor of $>$100 more sensitive and provide for the first time stringent experimental constraints on actinide nucleosynthesis in the last few hundred million years (see Figure 3). 

A simple steady-state scenario might represent a simplified assumption within our local ISM environment. Compared to the typical size of ISM substructures of  $\sim$50--100 pc (e.g. local bubble \cite{Man10,AL02,FBA06,BB14,Fer01}) and life-times of some 10 My, the crust sample probes, however, the equivalent of  $\sim$10 such cavities (the $^{244}$Pu life-time coupled with the spatial movement of the SS during the 24.5 My accumulation). Thus, we expect existing ISM inhomogeneities largely smeared out in our space- and time-integrated samples, confirming the significance of a ratio smaller than 1/100 between measured and expected $^{244}$Pu abundance. 

Further, we can relate our result to actinide nucleosynthesis during the recent SN history of the Local Bubble \cite{AL02,FBA06,BB14,BAF12,CA82} in which the SS is embedded now. ISM simulations suggest the LB was formed by  $\sim$14--20 SN explosions within the last 14 My \cite{AL02,FBA06,BB14} with the last one  $\sim$0.5 My BP \cite{AL02,BAF12}. In order to reproduce size and age of the LB, an intermediate density of 7 particles cm$^{-3}$ ($\sim$10 times the mean density of the local environment now) before the first SN explosion took place, is required \cite{BB14}. The mean mass density of the LB has since transformed to 0.005 particles cm$^{-3}$. The series of SNe explosions has generated the void inside the LB and has continuously pushed material into space forming an ISM shell. The SS is now placed inside the LB and thus has passed or passes a front of accumulated swept-up material including possible pre-existing $^{244}$Pu from nucleosynthesis events prior to the formation of the LB \cite{EFS96,FHE05}. 

\onecolumngrid
\begin{center}
\begin{figure}[ht]
\fboxsep=0pt
\includegraphics[width=16. cm]{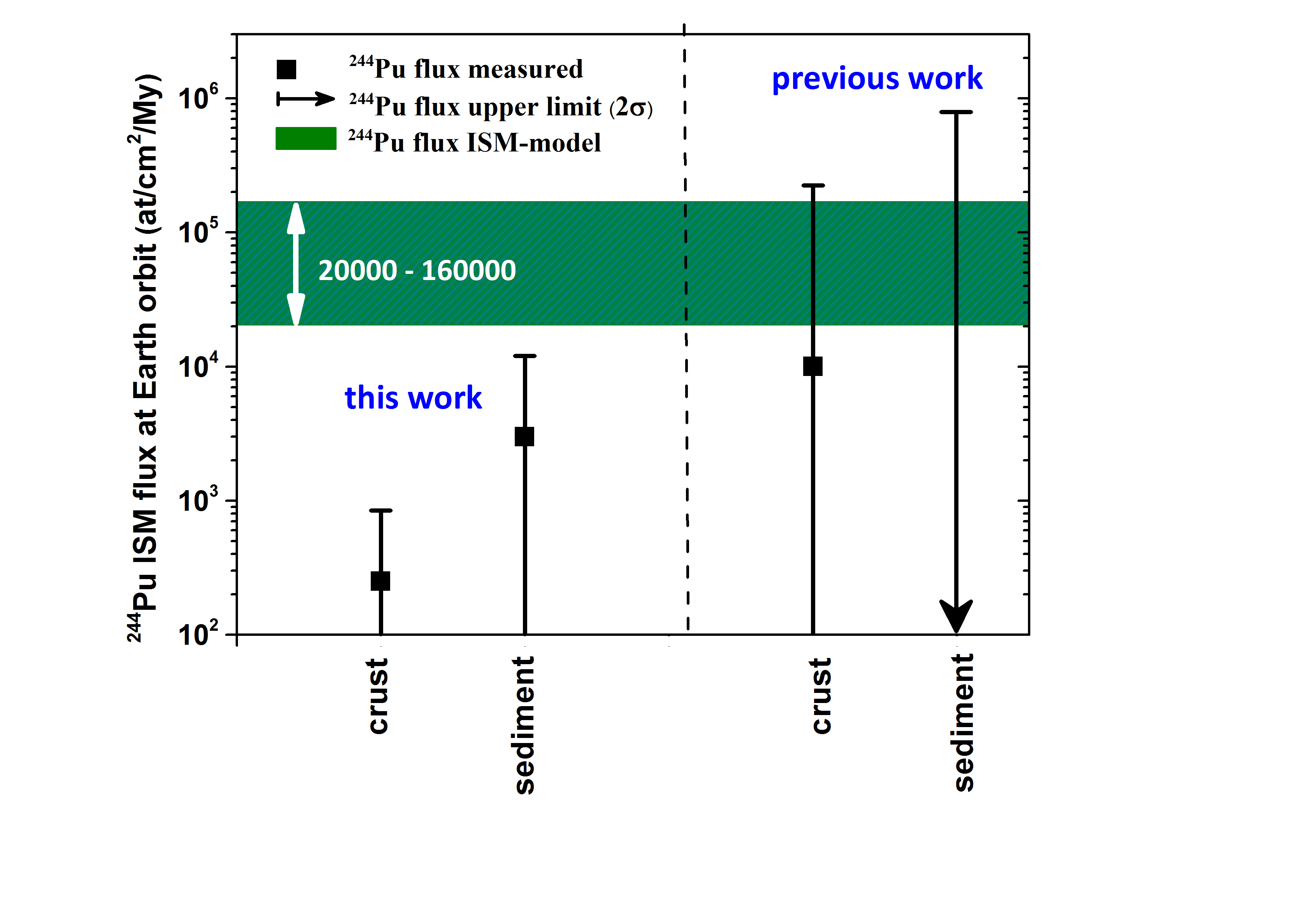}
\caption{ISM $^{244}$Pu flux at Earth orbit determined from the concentrations measured in a deep-sea crust and a deep-sea sediment sample (note the logarithmic scale). Our results are compared to previous measurements \cite{Wal04,Pau01} and to models of galactic chemical evolution \cite{WBG06,Cla11} assuming steady-state conditions and taking into account filtration of dust particles when entering the heliosphere \cite{Alt05}. The arrows and error bars represent upper levels (2$\sigma$, 95\% confidence levels) from the measurements. The green area indicates the data range deduced from the steady-state models. The crust data suggest a $^{244}$Pu flux which is a factor between 80 and 640 lower than inferred from the models.
\label{fig:fig3}}
\end{figure}
\end{center}
\twocolumngrid

We can distinguish 3 different scenarios for the recent LB history: (i) the SN activity transformed the local ISM from a dense to a low-density medium (LB), and pre-existing ISM material containing (steady-state) old $^{244}$Pu was swept-up and passed the SS \cite{EFS96,FFE14}; (ii) direct production of $^{244}$Pu in the 14--20 SNe and their expected traces left on Earth \cite{EFS96}; and (iii) independently, we can compare our data for $^{244}$Pu with recent AMS data of  $^{60}$Fe influx \cite{Kni04,Fit07}.

In a simple first order estimate for scenario (i), we assume that the swept-up material is distributed over a surface with a radius of 75 pc. Using the pre-LB density of 7 particles cm$^{-3}$ \cite{BB14} with 1\% of this ISM mass locked into dust, we calculate with our assumptions of Pu concentration in dust (Table II) and a dust penetration efficiency of  $\sim$6$\pm$3\% into the SS to Earth orbit, a $^{244}$Pu fluence from swept-up material of (0.4--3)$\times$10$^{6}$ $^{244}$Pu ats cm$^{-2}$ (see Table II).

Our experimental data give an flux of \tol{200}{800}{200} $^{244}$Pu ats cm$^{-2}$ My$^{-1}$ for the last 12 My (layers 2 and 3) at Earth orbit corresponding to a fluence of 2,300 ($<$12,000) $^{244}$Pu ats cm$^{-2}$ during this period. This experimental value for the fluence is a factor of  $\sim$170--1300 lower than the value calculated above (see Table II) assuming swept-up material of about half the diameter of the LB is moved across the SS. We deduce about the same discrepancy as found for a simple steady-state actinide production scenario. 

For the second LB-scenario, in a first order estimation, we take the SN-rate of 1.1--1.7 SNe/My within the LB \cite{BB14,BAF12} and a mean distance to the SS for these SN events of 100 pc. From our measured value of \tol{200}{800}{200} $^{244}$Pu ats cm$^{-2}$ My$^{-1}$ at Earth orbit (with 6\% penetration efficiency into 1 AU), this corresponds to  $\sim$3000 ($<$17,000) $^{244}$Pu ats cm$^{-2}$ My$^{-1}$ unfiltered ISM flux), spread over a surface area with radius of 100 pc and we deduce an average $^{244}$Pu yield per SN of (\tol{6 }{24}{6})$\times$10$^{-10}$  M$_{solar}$ for the last 12 My.

Finally for LB scenario (iii), Knie et al. \cite{Kni04} and Fitoussi et al. \cite{Fit07} measured a clear  $^{60}$Fe signal of possible SN origin  $\sim$2.2 My in the past in exactly the same crust material (237KD) as we have used in this work for the search for $^{244}$Pu (using a sample  $\sim$50 cm distant) (a SN origin for $^{60}$Fe is being questioned by some authors \cite{SL12,BSS07}; while several recent studies on $^{60}$Fe in deep ocean sediments \cite{Wal14,Lud14} and in lunar samples \cite{Fim14} confirm the results of Knie et.al. \cite{Kni04}). Thus we can directly compare the measured fluences of  $^{60}$Fe and $^{244}$Pu for the same event (using layer 2, 0.5--5 My). This fluence values can be converted into an atom ratio, that is independent on the SS penetration efficiency and we assume the same incorporation efficiency for Fe and Pu \cite{Wal14,Lud14,Fim14,FWW12}. 

We deduce a $^{244}$Pu/ $^{60}$Fe isotope ratio for this event of $<$6$\times$10$^{-5}$  (similarly, we obtain an upper limit from the sediment of $<$10$^{-4}$). Clearly, this ratio depends strongly on the type of explosive scenario. Literature values for this ratio are highly varying also due to large uncertainties in the r-process yields. 

Our experimental results indicate that SNe, at their standard rate of  $\sim$1--2/100 y in the Galaxy, did not contribute significantly to actinide nucleosynthesis for the past few hundred million years and actinide nucleosynthesis, as mapped through live $^{244}$Pu, seems to be very rare. Our data may be consistent with a predominant contribution of compact-object mergers which are 10$^{2}$  to 10$^{3}$ less frequent than core-collapse SNe \cite{Qia03}. A recent observation indicates indeed that such mergers may be sites of significant production of heavy r-process elements \cite{AST04,GA01}. Our experimental work is also in line with observations of low metallicity stars \cite{CS06,JF13} indicating a splitting into a rare and a more frequent r-process scenario allowing an independent evolution of the r-process elements Eu/Th over time \cite{AGT07,Thi11}. 

In addition, we must conclude from our findings that, given the presence of short-lived actinide $^{244}$Pu (and  $^{247}$Cm) in the ESS, it must have been subject to a rare heavy r-process nucleosynthesis event shortly before SS formation.

\vspace*{1mm}
{\bf Acknowledgements}: We thank the following organizations for supporting this work: part of this work was funded by the Austrian Science Fund (FWF): Project No. AP20434 and AI00428 (FWF \& CoDustMas, Eurogenesis via ESF). We thank A. Lueckge and M. Wiedicke, Bundesanstalt f{\"u}r Geowissenschaften und Rohstoffe, Stilleweg 2, D-30655 Hannover, Germany for providing us the VA13 crust sample; and D. Lal and W. Smith (Scripps Geological Collections, US) for locating and providing us with the deep-sea sediment samples (TRIP core).

\vspace*{1mm}
{\bf Author Contributions}: All authors were involved in the conception and planning of the project and in the writing of the paper. A.W. performed the data analysis and wrote the main paper together with M.P., and all authors discussed the results and commented on the manuscript. K.K., T.F. and G.K. organised the crust sample; M.P. provided the sediment sample. K.K., F.Q. were primarily responsible for sample preparation of the crust; A.O and C.F. for the sediment. P.S., A.W. and K.K. performed the AMS measurements. 
\vspace*{1mm}

{\bf Author Information}: The authors declare no competing financial interests. 
Supplementary information accompanies this paper on www.nature.com.
Correspondence and requests for materials should be addressed to A.W. (anton.wallner@anu.edu.au).

\section{Methods: \label{sec:4}}

\subsection{Details on the chemistry of the crust samples}
Quantitative extraction of Pu was required from the 2.3 kg-sized crust sample. The sample potentially contained some 10$^{6}$-10$^{7}$ atoms of $^{244}$Pu which corresponds to an atom-concentration of (0.5--5)$\times$10$^{-19}$ relative to the bulk material. No stable isobar to $^{244}$Pu exists in nature and molecular interference in the measurements is excluded. The 12 individual pieces had masses between 30 and 360 g. 10 samples were measured by AMS. Four different vertical layers represent different time periods in the past, while three different horizontal sections were chosen to identify possible lateral variations (B, C and D). 

The individual parts of the crust were dissolved in aqua regia and H$_{2}$O$_{2}$, and a spike of  $\sim$3 $\times$10$^{8}$ atoms of a $^{236}$Pu reference material was added to the leached solutions. After removal of the undissolved SiO$_{2}$ fractions, the solutions were brought to dryness and successively redissolved in concentrated HNO$_{3}$ and H$_{2}$O$_{2}$. At this stage, the sample solutions contain the actinides, but also the dominant fraction of the matrix elements of the crust, in particular Mn ($\sim$14 to 28\%) and Fe ($\sim$16 to 28\%) \cite{FS80}. 

In order to separate the actinides from the Mn and Fe fraction, a pre-concentration step involving the selective co-precipitation of the actinides with CaC$_{2}$O$_{4}$ at pH  $\sim$1.7 was performed. After centrifugation, the precipitated CaC$_{2}$O$_{4}$ was converted to CaCO$_{3}$ in a muffle furnace at a temperature of 450$^{\circ}\mathrm{C}$ for several days. The CaCO$_{3}$ was dissolved in 7.2 M HNO$_{3}$ and the oxidation state of Pu was adjusted quantitatively to (IV)Pu by the addition of NaNO$_{2}$. These solutions were then loaded onto pre-conditioned anion-exchange columns containing Dowex 1$\times$8, from which after the separation of the Ca, Am, Cm and the Th fractions, Pu was eluted by reduction to (III)Pu with a solution of HI. In order to purify the obtained Pu fraction, two additional successive anion-exchange separations similar to the one described above were performed on the eluted Pu solutions. The Pu fractions were further purified from the organic residues by fuming with HNO$_{3}$ and H$_{2}$O$_{2}$. Successively an Fe(OH)$_{3}$ co-precipitation of Pu was performed in 1M HCl by adding 2 mg of Fe powder. After centrifugation and drying of the precipitate, the Fe(OH)$_{3}$ was converted to iron oxide by combustion in a muffle furnace at 800$^{\circ}\mathrm{C}$ for 4 hours. The plutonium oxide embedded in a matrix of iron oxide was then mixed with 2 mg of high-purity Ag powder and pressed in the samples holders suitable for the subsequent AMS measurement.

Owing to the massive matrix component of the crust \cite{FS80} for some samples a low chemical yield was observed (in the AMS measurements via the $^{236}$Pu spike). In these cases the procedure for Pu extraction was repeated, i.e. the solutions containing the Mn and Fe fraction underwent again a CaC$_{2}$O$_{4}$ co-precipitation procedure and the resulting actinide fractions were mixed with the rest of the remaining fractions originating from the first three-fold anion-exchange separation. These solutions underwent a chromatographic column separation employing Tru-resinTM in 5M HCl, from which, after the elution of the Ca, Am and Cm and the Th fractions, Pu was stripped out with a solution of 0.03M H$_{2}$C$_{2}$O$_{4}$ in 0.5M HCl. 

\subsection{Chemical processing of the TRIP deep-sea sediment}

Similarly, Pu was extracted from two deep-sea sediment samples of 43 and 58 g mass provided by the Scripps Oceanographic Institute, University of California at San Diego. It was a piston core (7P), extracted during the TRIPOD expedition (1966) as part of the Deep-Sea Drilling Project (DSDP) at location $17^\circ$ $30^\prime$ N, $113^\circ$ $00^\prime$W (Pacific Ocean) at 3763 m water depth. Two main sections were sampled with sediment depths 0--80 cm and 80--230 cm, from which the top 3 cm (containing the anthropogenic Pu) were removed. One quarter of the total cross section throughout the  $\sim$230 cm length of the core was used in this study (4.9 cm$^{2}$). The samples, bagged in sealed polyethylene, were chemically processed at the Hebrew University, Jerusalem. 

The physical and chemical processing of the sediments was the following: the processing involved brief milling and calcination of the sample, alkali fusion of the sediment using NaOH at 750$^{\circ}\mathrm{C}$, liquid-phase extraction of Fe and other main elements and liquid ion-chromatography to extract the Pu fraction. Prior to the alkali fusion steps, an isotope $^{242}$Pu marker and a chemical  $^{230}$Th marker were added to the sediment. $^{242}$Pu was used to monitor the efficiency of Pu detection by measuring the $^{242}$Pu content of the final AMS sample (analogous to the $^{236}$Pu spike in the crust samples) while  $^{230}$Th served as additional indicator of the chemical efficiency of actinide extraction by measuring the alpha activity of an electroplated deposition prepared from a separate fraction. The final AMS samples were obtained by co-precipitation of the Pu fraction with Fe in an ammonia solution, centrifugation and ignition to obtain a Fe$_{2}$O$_{3}$ matrix containing the Pu marker and traces. Finally, 2 mg of high-purity Ag powder was added and the powder pressed in the samples holders suitable for the subsequent AMS measurement.

\subsection{AMS-measurement procedure}

We have applied the most sensitive technique, AMS \cite{Ste13,Syn13,Kut13} to detect minute amounts of $^{244}$Pu. This technique provides the complete suppression of any interfering background, e.g. molecules of the same mass, by the stripping process in the terminal of a tandem accelerator, which is crucial for such experiments where only a few counts are expected. The $^{244}$Pu measurements were performed at the VERA (Vienna Environmental Research Accelerator) facility at the University of Vienna \cite{Ste13,Wal10,Wal14b}. This setup has been optimized for high-measurement efficiency and offered an exceptional selectivity.

The individual crust samples were spiked with a well-known amount of $^{236}$Pu atoms ($^{242}$Pu for the sediment). $^{244}$Pu and $^{239}$Pu measurements were performed relative to the $^{236}$Pu ($^{242}$Pu) spike, i.e. the total efficiency and the chemical yield of Pu (when compared to the theoretical measurement efficiency) was monitored with the $^{236}$Pu ($^{242}$Pu) spike, that was counted in short time intervals before and after the $^{244}$Pu runs in the AMS measurements. The chemical yield was varying between 5\% and 70\% largely depending on the sample matrix.

In a sputter source the Fe/Ag matrix, containing the Pu atoms, was bombarded with Cs ions and negative PuO ions were extracted, energy and mass analysed before injection into the tandem accelerator. The negative ions were accelerated to the 3-MV tandem terminal and stripped there to positive ions in the gas stripper (O$_{2}$). $^{244}$Pu$^{5+}$-ions were accelerated to a final energy of 18 MeV and selected with a second analyzing magnet. They then had to pass an additional energy and another mass filter (electrostatic and magnetic dipoles respectively) and were finally counted in an energy-sensitive particle detector. 

The system was optimized with a $^{238}$U pilot beam and monitored during a measurement with reference samples containing $^{242}$Pu. The measurement setup for $^{244}$Pu and $^{236}$Pu counting was scaled from the tuning setup. At the end of a measurement series, reference samples containing a well-known isotope ratio of $^{244}$Pu/$^{242}$Pu were measured. The measured ratios reproduced the nominal values within 4\% and confirmed the validity of scaling between the different masses in the measurement.

This setup suppresses adjacent masses (e.g. $^{238}$U from $^{239}$Pu) by 8 to 9 orders of magnitude. This suppression factor was sufficient for $^{244}$Pu counting as no interference from neighbouring masses is expected for $^{244}$Pu (and additional isotopic suppression, e.g. via time-of-flight identification, would have been at the cost of lower particle transmission). However, $^{238}$U was abundant at levels of  $\sim$10 ppm in the crust and U separation from Pu in the chemical preparation of these samples was not 100\%. Thus, the detector events registered for $^{239}$Pu counting, lower by two to three orders of magnitude compared to modern samples, are attributed to leaking $^{238}$U atoms injected as $^{238}$UOH$^{-}$ ions together with $^{239}$PuO$^{-}$, thus mimicking $^{239}$Pu. To summarize, our detector event rate for $^{239}$Pu suggests no significant anthropogenic $^{244}$Pu contamination.

The measurement procedure was a sequence of alternating counting periods of $^{236}$Pu, $^{239}$Pu, $^{244}$Pu and again $^{236}$Pu. All samples were repeatedly measured until they were completely exhausted. The sputtering time per sample was between 5 and 20 hours. Three measurement series were required to fully consume all the samples. The overall yields for the 12 crust samples were between 0.06$\times$10$^{-4}$ and 1.54$\times$10$^{-4}$; i.e. one $^{244}$Pu detector event would represent correspondingly between 6500 and 1.7 $\times$10$^{5}$ $^{244}$Pu atoms in the analysed sample. A similar procedure was followed for the sediment samples where $^{242}$Pu was used as a spike.

Due to the low number of expected detector events, the machine and measurement background was carefully monitored with samples of pure Fe and Ag powder. They were sputtered identical to the samples containing the crust fractions. In addition, one crust sample (X), the lowest layer of the crust material, was of hydrothermal origin where no extraterrestrial $^{244}$Pu could accumulate. This sample was chemically prepared and measured in the same way as the other crust samples and served as a process blank for potential chemistry and machine background.

\subsection{Incorporation efficiency of Pu into the deep-sea crust}

The incorporation efficiency of bomb-produced Pu into the crust was determined from the anthropogenic $^{239}$Pu content in the top layer of the crust and deep-sea sediment data (details are given in Table 3). The average over 18 sediment cores from the Pacific measured between 1974 and 1979 gave a $^{239,240}$Pu sediment inventory of 2.15 Bq m$^{-2}$ \cite{LA83}. When compared to the well-known surface activity of 2.8 mCi km$^{-2}$ (10$^{4}$ Bq m$^{-2}$) \cite{BNL80}, at the time of sampling the crust in 1976, 2.1\% of anthropogenic Pu from the bomb-tests was incorporated into deep-sea sediments with sediments having an incorporation efficiency of 100\% (this compares well with the ratio of the time that had passed since the peak in atmospheric bomb-testing (15 y) and the mean residence time of Pu in the ocean of  $\sim$440 y; i.e. 3.4\%). 

Taking the total anthropogenic Pu-inventory at the location of the crust (between 60 and 78 Bq m$^{-2}$ and a $^{240}$Pu/$^{239}$Pu atom-ratio of 0.20) \cite{Wal04,Pau01} and the fraction of 2.1\% measured in sediments, we calculate 8.2 $\times$10$^{7}$ $^{239}$Pu atoms cm$^{-2}$ have reached the crust in the year 1976. We measured from the three crust subsamples from the top layer a $^{239}$Pu surface-density of (1.76 $\pm$0.44) $\times$10$^{7}$ at cm$^{-2}$, thus we deduce an incorporation efficiency into the crust of (21 $\pm$5)\%. We assume that the ISM-Pu is incorporated like the bomb produced Pu.

\newcommand{\noopsort}[1]{} \newcommand{\printfirst}[2]{#1}
\newcommand{\singleletter}[1]{#1} \newcommand{\swithchargs}[2]{#2#1}

\end{document}